\UseRawInputEncoding
\documentclass{article}

\PassOptionsToPackage{numbers, compress}{natbib}

\usepackage[preprint]{neurips_2024}

\usepackage{natbib}
\usepackage[utf8]{inputenc} 
\usepackage[T1]{fontenc}    
\usepackage{hyperref}       
\usepackage{url}            
\usepackage{booktabs}       
\usepackage{amsfonts}       
\usepackage{nicefrac}       
\usepackage{microtype}      
\usepackage{xcolor}         
\usepackage{amsmath}
\usepackage{graphicx}
\usepackage{natbib}

\title{Narrative Information Theory}

\author{%
  Lion Schulz$^{1}$ \quad
  Miguel Patrício$^{2}$ \quad
  Daan Odijk$^{2}$ \\[.5em]
  $^1$Bertelsmann \quad $^2$RTL Nederland \\[.5em]\
 \texttt{lion.schulz@bertelsmann.de}
}

\begin{document}
\setcitestyle{numbers}

\maketitle
\begin{abstract}
We propose an information-theoretic framework to measure narratives, providing a formalism to understand pivotal moments, cliffhangers, and plot twists. This approach offers creatives and AI researchers tools to analyse and benchmark human- and AI-created stories. We illustrate our method in TV shows, showing its ability to quantify narrative complexity and emotional dynamics across genres. We discuss applications in media and in human-in-the-loop generative AI storytelling.
\end{abstract}

\section{Introduction}

As AI systems begin to tell their own stories \citep{sun_language_2023, chu_ai_2017}, it becomes crucial to develop formal methods for understanding and evaluating the content they produce \citep{piper_narrative_2021, del_vecchio_data_2021, vishnubhotla_emotion_2024, sap_quantifying_2022}. We introduce a general information-theoretic framework to measure narratives, capturing key storytelling elements like novelty and surprise \citep{piper_modeling_2023, murdock_exploration_2017, barron_individuals_2018}.  
This work provides tools for creatives and for applied machine learning scientists to analyse narrative structures and for GenAI researchers to benchmark stories told by machines. It thereby also offers a foundation for developing human-in-the-loop AI systems assisting in narrative creation while maintaining the nuanced human understanding of stories \citep{pizzo_interactive_2023}.

Narratives represent a key higher-level representation of how a story gets told \citep{piper_narrative_2021} and are crucial to how we understand worlds, both fictional and real. Previous machine learning work on detecting narratives has focussed on specific content and modalities \citep{ teodorescu_evaluating_2023, agarwal_shapes_2022, kim_frowning_2019, hipson_emotion_2021, crijns_multimodal_2020, elkins_shapes_2022}.  We build on this by introducing a general information-theoretic narrative framework that is agnostic to content and modality and lets us intuitively capture crucial  storytelling devices, like novelty and surprise -- which have thus far seen less attention in the study of narratives. We illustrate our measures in a corpus of over 3000 minutes of TV shows, one of the most popular forms of contemporary storytelling. 

\begin{figure}[h!]
  \centering
  \includegraphics[scale=.16]{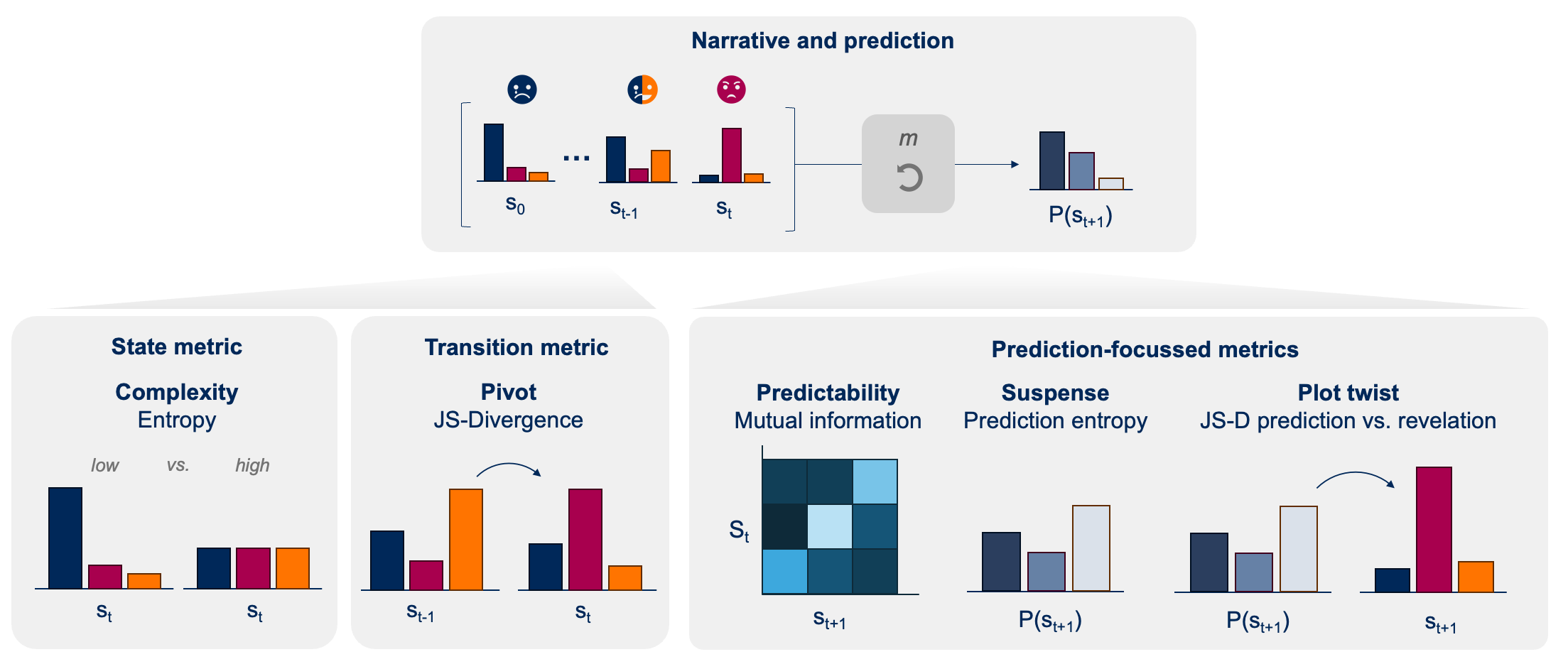}
  \caption{Overview -- information-theoretic measures of narratives.}
  \label{fig:overview}
\end{figure}

\section{Framework and Results}

\subsection{States, complexity and pivots}

We can intuitively grasp what is going on in a story: a happy scene follows a sad scene, cliffhangers leave us unsure about what will happen next. Here, we show how we can capture such dynamics using information theory. To construct our formal framework, we begin by decomposing a story into basic building blocks that we call states at each timepoint $t$ in the narrative, $s_t$ (see Fig. \ref{fig:overview} for an overview). We are agnostic about the nature of this state. It might contain the setting of a scence, the emotions of the characters, or be based on a more complex models of the story. 

For illustration, we applied our framework to a corpus of TV shows where we defined states as distributions of emotions inferred from actor faces (see appendix).  However, we note that our framework is agnostic to the modality of the story told and so could also be applied to written text such as books or audio. Fig \ref{fig:results}A shows an example trajectory of states across an episode of a drama TV show set in a crime setting. The episode starts out on a sad note, transitions into a more neutral phase, and then ends again on two sad beats.

Our first information theoretic measure to better understand $s_t$ is the entropy of this state which we can understand as the complexity \citep{nath_relating_2023} of a narrative state:

\begin{figure}[t!]
  \centering
  \includegraphics[scale=.237]{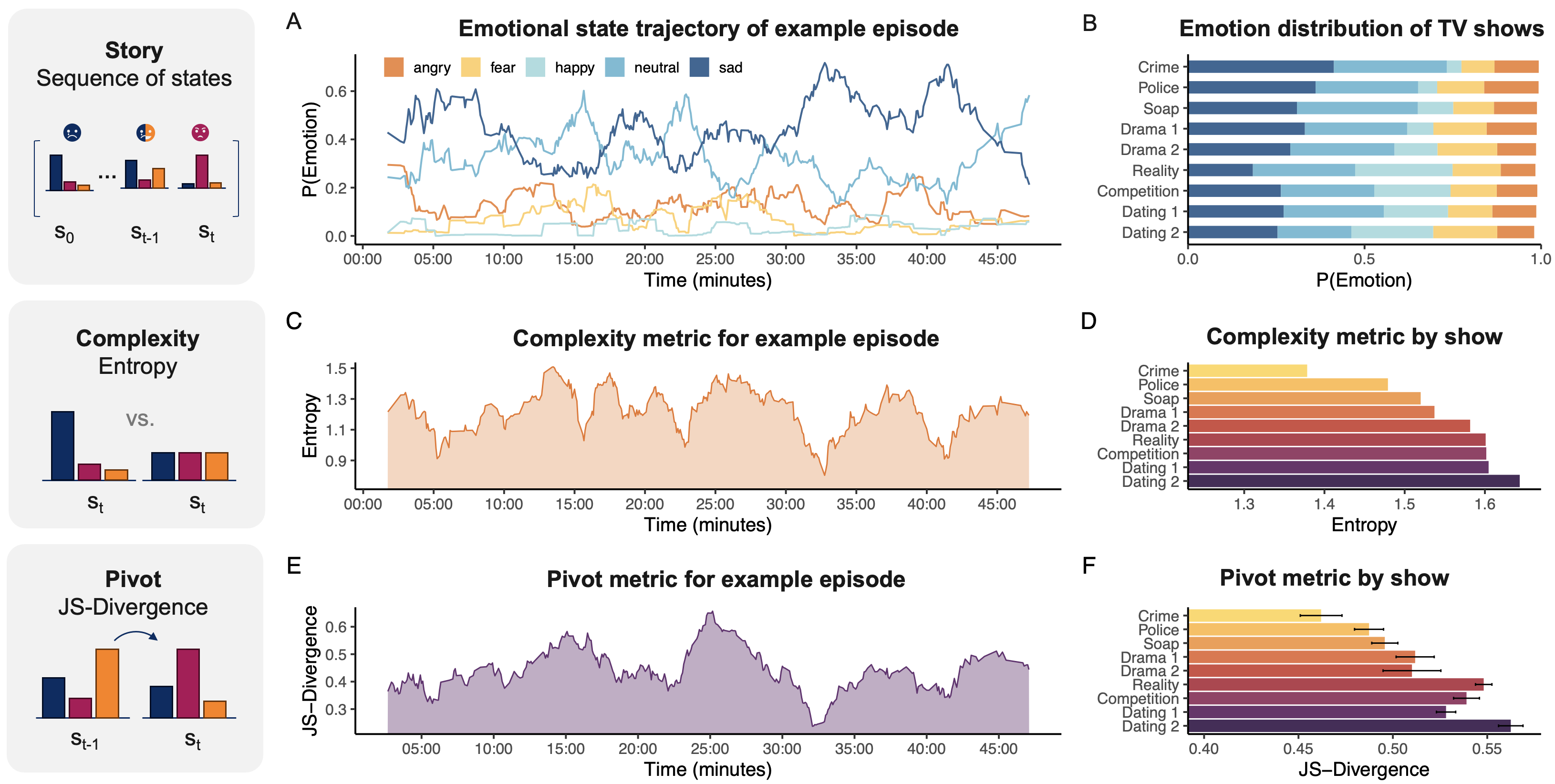}
  \caption{Results -- emotions, as well as the complexity and pivot metrics in an example episode (crime thriller) and across different shows (see appendix for details).} 
  \label{fig:results}
\end{figure} 

\begin{equation}
    \textbf{Complexity} = \text{H}(s_t)
\end{equation}

In our example emotion case, the entropy of a scene will be lower when a distribution is dominated by one emotion (everyone is happy) but higher when there is a mixture of emotions. We show this in Fig \ref{fig:results}C where we plot the state by state entropy for our example episode. The peaks in sadness in the episode's finale are characterized by valleys in entropy, whereas the middle of the show is characterized by comparatively higher entropy because of its mix of emotions.

Our entropy measure can operate at different resolutions: For example, books are made up of chapters and paragraphs, and movies can be split into individual scenes. In turn, the complexity of these entire (sub-)stories can be expressed by the entropy of the average of the states. Consequently, a book that contains mostly sad scenes will have generally low entropy, whereas a movie with a mix of emotions will have high entropy. 

We apply this analysis at the level of entire TV shows, analysing our corpus of over 3000 minutes of video across genres as broad as crime thrillers, historical dramas, and reality TV (see appendix for details). Our analysis revealed distinct patterns across genres (see Fig. \ref{fig:results}D). Reality formats showed higher entropy, indicating a broader mix of emotions (see the underlying emotion distributions in Fig. \ref{fig:results}B). In contrast dramas/thrillers had lower entropy, focussing on specific emotional tones. 

We can extend this approach to story dynamics, or transition between states \citep{piper_modeling_2023, barron_individuals_2018}. A character might start out happy and then be sad. We describe such pivots as the Jensen-Shannon-divergence (JSD) between two states: 

\begin{equation}
    \textbf{Pivot} = \text{JSD}(s_t \mid \mid  s_{t-1})
\end{equation}

Strong shifts in the state, for example a move from a shot full of mainly happy character to one dominated by angry characters would result in a high divergence. In contrast, two happy scenes following each other lead to a low divergence. We plot this metric for our example TV episode in Fig. \ref{fig:results}E. Peaks indicate moments of significant shifts in emotional state (e.g. in our example at around 25:00) - essentially a story beat. Taking the liberty of more poetic language, we can interpret this curve as the heartbeat of the story.\footnote{We here choose JSD over Kullback-Leibler Divergence because of its symmetry and because its boundedness makes it more suitable for usage as a metric \citep{vrijenhoek_radio_2024}. We observe similar qualitative results with KLD, see appendix.}

Our pivot measure also revealed genre-specific patterns (see Fig. \ref{fig:results}F). Reality and dating shows displayed higher JSD, indicating frequent and stronger emotional shifts – essentially an "emotional rollercoaster" metric.  This is compared to genres that rely on comparitively more gradual changes like dramas and thrillers - often "slower burns".
These results demonstrate our framework's ability to quantify narrative structures and emotional dynamics, providing a basis for comparing storytelling techniques across different genres and between human-created and AI-generated content.

\subsection{Prediction-focussed metrics: suspense and plot twists}

Audiences do not just observe what happens at time $t$ but also wonder what happens next (see Fig. \ref{fig:overview}). We consider them to predict a next story state based on the story so far. We write this prediction as $P(s_{t+1}|S_t)$, essentially the output of a (generative) model $m$ that takes $S_t = \{s_t, s_{t-1}, ..., s_0\}$ as its input \citep{barron_individuals_2018}. In this paper, we outline theoretically how we can understand these predictions and reserve analysis and the accompanying model development for future work.  We remain agnostic towards the underlying model but note how it is essentially a sequence-to-sequence prediction problem.

First, we can ask how predictable a story is via the mutual information between the history and a future, realised state. High mutual information means a show is predictable - that is, if we know the history of the show, we can make better predictions about its future:

\begin{equation}
    \textbf{Predictability} = I(s_{t+1};S_t)
\end{equation}

A key question for an audience is how much it remains in the dark about what will happen next. We can capture this uncertainty in information-theoretic terms by computing the entropy over the predicted distribution: 

\begin{equation}
    \textbf{Suspense} = \text{H}(P(s_{t+1}|S_t))
\end{equation}

In simple terms, this entropy measure captures how confidently the audience can predict what will happen next. If it is low, a film viewer would (feel like they) clearly know what will happen in the next scene. In turn, if the entropy is high, we are left wondering what will happen next. Shows with cliffhangers would see a spike in this entropy towards the end of an episode whereas closed endings should have this entropy measure decline in the final chapters. 

When predictions meet reality, we can again quantify an audience's reaction by computing the JSD between the prediction and the revealed reality of the story: 

\begin{equation}
    \textbf{Plot twist} = \text{JSD}(P(s_{t+1})\mid\mid s_{t+1})
\end{equation}

This measure can capture key intuitions about a story: A high divergence will signify an unexpected plot twist. Low divergence will mean that a story treads along as expected. We note how these prediction-based measures are orthogonal to the measures introduced at the outset. Take for example our two JSD measures: On the one hand, changes in the state might be entirely predictable. On the other, the  absence of a change might be suprising if it violates predictions. For example, imagine a character receiving good news but staying sad.

\section{Discussion}

Our work introduces an information-theoretic framework to capture narratives. We showed how core principles from information theory provide a formal language for understanding story dynamics, and applied them to a real-world data set. Beyond these measures operating on the state of a narrative, we formally introduced metrics for predicted future courses of a storyline. 

Our framework provides a foundation for several key areas of interest at the intersection of creativity and AI: For example, our measures offer quantitative metrics for assessing the complexity, unpredictability, and plot twists in stories. The genre-specific patterns observed in TV shows could serve as baselines for evaluating AI-generated stories in different styles. They may also help distinguish between human and AI-generated narratives, potentially revealing characteristic patterns in AI stories. Our metrics may also assists in identifying systemic biases in narratives across genres or cultures -- ensuring AI-generated content reflects diverse storytelling traditions. Finally, our framework might assist human and AI co-creators, suggesting plot developments or highlighting areas that may need more tension or surprise to maintain audience engagement. 

For more classical ML applications, our metrics can serve as metadata for viewership analyses and recommendations \citep{zhang_frontiers_2020},  e.g. differentiating people's interest for more or less (emotionally) complex movies. One may also investigate how likely people are to watch a next episode of a TV series based on the quantified strength of a cliffhanger.  Our metrics could also help both human and AI editors identify crucial and relevant moments, accelerating summary or trailer generation \citep{bretti_find_2024}. For example, in summarisation, practicioners highlight pivotal moments. In turn, trailers or previews might want to avoid revealing plot twists.

Future work should focus on these applications and extend the framework to other modalities like books \citep{piper_modeling_2023, toubia_how_2021} or even music \citep{cohen_information_1962}. For prediction-focused measures, applying (generative) models to capture human intuitions about stories will be a key challenge. An LLM's next token distribution might be a starting point. Creative professionals' attitudes towards such quantitative narrative analysis tools should guide further development.
In conclusion, we offer a bridge between the creative intuitions of human storytellers and the analytical capabilities of ML, contributing to the ongoing dialogue about how AI can augment and enhance human creativity rather than replace it.

\begin{ack}
We would like to thank Franziska Brändle, Peter Dayan, Dante Di Loreto, Tom Hoffman, Kit Thwaite and the Penguin Random House UK data science team for comments and discussions. We are particularly grateful to the RTL NL data science team for support and discussions, especially Mateo Gutierrez Granda, Iskaj Janssen, Prajakta Shouche and Ivan Yovchev. The authors are employees of Bertelsmann SE \& Co. KGaA (LS) and RTL Nederland BV (MP, DO).
\end{ack}

\bibliographystyle{ieeetr}
\bibliography{manual_refs}


\appendix

\section{Appendix / supplemental material}

\subsection{Dataset}

For the analysis that is illustrated in Figure \ref{fig:overview}, we had access to an internal dataset from a large European network. This dataset encompassed over 3000 minutes of video and represents a broad overview of the landscape of the most popular TV formats. It contained:

\begin{itemize}
    \item 16 episodes of a long-running daily soap focusing on the intertwined lives of residents of a small community (abbreviated as "Soap" in our plots). 
    \item 12 episodes of a dating format where couples test the strength of their relationships by living with attractive singles on a tropical island ("Dating 2").
    \item 11 episodes of survival competition format where contestants are stranded on a remote island and must work together to overcome challenges ("Competition").
    \item 8 episodes of a reality series where single individuals run a bed and breakfast while searching for love ("Dating 1").
    \item 8 episodes of a gritty crime drama exploring the lives and conflicts within a criminal underworld ("Crime").
    \item 6 episodes of a thriller series that delves into the secretive world of espionage and undercover operations ("Police").
    \item 5 episodes of a crime drama based on true events, focusing on the complexities of family betrayal within a notorious crime family ("Drama 1").
    \item 4 episodes of a biographical drama centered on the life and challenges of a prominent public figure ("Drama 2").
    \item 3 episodes of a reality show following the everyday lives and humorous antics of a well-known family ("Reality").
\end{itemize}

\subsection{Analysis pipeline}

To analyse the individual episodes, we applied a machine learning pipeline. In short, this pipelines adhered to the following steps. We first subsampled the videos to extract a frame every 5 seconds. On these frames, we then applied face detection. On the detected faces, we applied emotion detection using the deepface library. This library outputs a distribution over 7 emotions ("angry", "disgust", "fear", "happy", "sad", "surprise", "neutral"). 

For our analysis, we then applied the following steps: We first averaged the distributions per frame (here using a rolling average over 20 extracted frames that contain a face). Fig. \ref{fig:overview}A plots these averaged values for the distributions of an episode of the "crime" series. We plot the entropy of these averaged states in Fig. \ref{fig:overview}C. Fig \ref{fig:overview}D in turn shows the JSD between raw states, again averaged over 20 frames with faces. We note that we only plot the most prominent 5 emotions in Fig. \ref{fig:overview}A and B for readability - which make up more than 95 \% of the emotions in the shows. Entropies and JSD in turn were computed on the seven original emotions.

To compute entropies per show, we averaged all states across episodes to form an average emotion distribution per show (Fig. \ref{fig:overview}B), and computed the entropy based on this single distribution per show (Fig. \ref{fig:overview}D). We used this single distribution because it allowed us the best high-level overview of the emotions contained in a show. In turn, to compute average values for the pivot-metric, we first averaged the JSD per episode, and then plotted the per-show means of these episode-level JSD's  and the accompanying standard error of the mean in Fig \ref{fig:overview} F.

\begin{figure}[t!]
  \centering
  \includegraphics[scale=.20]{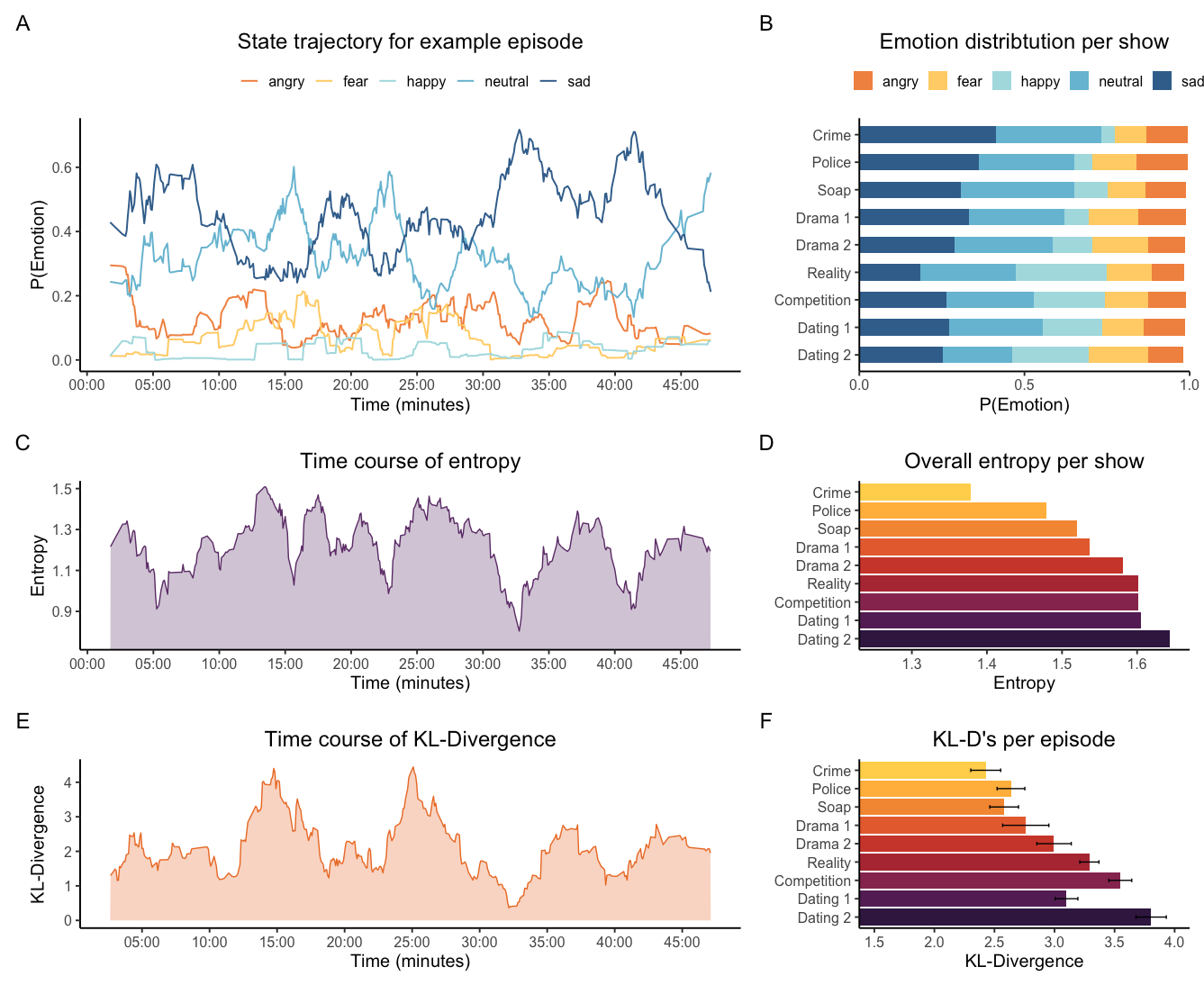}
  \caption{Supplementary figure: KL-Divergences for example show and trajectories (left) and summary statistics (right). Note the overlapping qualitative patterns with JS-D.}
  \label{fig:KLD}
\end{figure}

\subsection{Choice of divergence measure}

We here choose JSD over Kullback-Leibler Divergence because of its symmetry and because its boundedness makes it more suitable for usage as a metric \citep{vrijenhoek_radio_2024}. We observe similar qualitative results with KLD, see Fig. \ref{fig:KLD}.

\end{document}